\documentclass{jltp}
\input psbox.tex
\def\beq{\begin{equation}}
\def\eeq{\end{equation}}
\def\no{\noindent}
\def\z{\zeta}
\def\a{\alpha}
\def\L{\Lambda}
\def\s{\sigma}

\def\cos{{\rm cos}}

\def\det{{\rm det}}

\def\vphi{\varphi}

\def\bchi{{\bar\chi}}
\def\bvphi{{\bar\varphi}}
\def\dphi{{\delta\varphi}}
\def\dchi{{\delta\chi}}
\def\bpsi{{\bar\psi}}
\def\brho{{\bar\rho}}

\def\ph{\Phi}
\def\bph{{\bar \phi}}
\def\Ph{\phi}

\def\bpsi{{\bar \psi}}

\def\eps{\epsilon}

\def\a{\alpha}
\def\cos{{\rm cos}}

\def\om{\omega}
\def\s{\sigma}

\def\om{\omega}

\def\Dphi{\delta\phi}

\def\a{\alpha}
\def\s{\sigma}
\def\cG{{\cal G}}

\def\cos{{\rm cos}}

\def\det{{\rm det}}

\def\vphi{\varphi}

\def\bchi{{\bar\chi}}
\def\bvphi{{\bar\varphi}}
\def\dphi{{\delta\varphi}}
\def\dchi{{\delta\chi}}
\def\bpsi{{\bar\psi}}
\def\brho{{\bar\rho}}

\title{Interacting Bose Gas in an Optical Lattice\thanks{Dedicated to 
Peter W\"olfle on the occasion of his 60th birthday.}}

\author{K. Ziegler}

\address{Institut f\"ur Physik, Universit\"at Augsburg, Germany}

\runninghead{K. Ziegler}{Interacting Bose Gas in an Optical Lattice}

\begin{document}

\maketitle
\begin{abstract}
A grand canonical system of hard-core bosons in an optical lattice 
is considered. The bosons can occupy randomly $N$ equivalent states
at each lattice site. The limit $N\to\infty$ is solved exactly
in terms of a saddle-point integration, representing a weakly-interacting
Bose gas. In the limit  $N\to\infty$ there is only a condensate
if the fugacity of the Bose gas is larger than 1. Corrections in $1/N$
increase the total density of bosons but suppress the condensate. This
indicates a depletion of the condensate
due to increasing interaction at finite values of $N$.

PACS numbers: 03.75.Fi, 05.30.Jp, 32.80.Pj, 67.40.-w
\end{abstract}


\section{INTRODUCTION}

Recently developed systems of trapped ultra-cold gases open a wide new field
for the study of many-particle
physics.\cite{anderson}
This includes the investigation of Bose-Einstein condensation of
interacting Bose particles\cite{feynman,huang},
leading to macroscopic quantum states or ``matter waves''. Among the most
interesting experiments there
is the trapping of bosons in so-called optical 
lattices\cite{jaksch,oosten,chiofalo}
that are created by intersecting Laser fields. Typical lattice
constant of optical lattices are $a_l=250 ... 500$nm.
They exhibit unsual properties due to the interference of
macroscopic quantum states. The interaction between the atoms is
characterized by the ratio of the
scattering length $a$ and the typical particle distance $n^{-1/3}$ as
$a n^{1/3}$, where $n$ is density of particles. 
In most of the experiments we have $ a n^{1/3}<10^{-2}$ such that these
systems are dilute\cite{anderson}.
Interaction plays only a weak role in these systems such that the
Gross-Pitaevskii approach is sufficient\cite{ginzburg,baym}.
After the experimental discovery of Feshbach resonances in the
trapped Bose gases
it became possible to tune the inter-atomic interaction 
with an applied magnetic field  over a wide range of 
interactions\cite{courteille,roberts,cornish}.
The scattering length for $^{85}$Rb
atoms is $a\approx 5$nm but near Feshbach resonances it is up to 
500nm\cite{cornish}.
In this case one can study an optical lattice in which
the lattice constant
(determined by the wave length of the applied Laser field) is equal to the
scattering length. Then
at most one boson can be found in the minima of the optical lattice.  

From the theoretical point of view this field
requires the study of models with strongly
interacting particles. 
The so-called Bose-Hubbard model has been
used\cite{jaksch,oosten} to decribe an on-site interaction with
coupling strength $U$. A mean-field calculation\cite{oosten} has
shown that the condensate vanishes in the strong-coupling limit
$U\to\infty$, indicating that a strong interaction among the
bosons has a destructive effect on the condensate.
On the other hand, from the experiments it is believed that
the interaction can still be considered as hard-core,
characterized only by the
scattering length $a$ but including many-particle collisions.  

In the following we will tune the interaction by going from the weakly
interacting (dilute) to the strongly interacting hard-core Bose gas.
This will be achieved by the assumption that each lattice site can
accommodate $N$ bosons with the same energy\cite{ziegler2}. 
Bosons can randomly change between these states and tunnel 
between neighboring lattice sites from/to
any of these $N$ bosonic states with the same rate. 
Interaction exists only between bosons at the same state.
The regime $N\gg1$ describes a dilute Bose gas whereas $N=1$ corresponds to a 
hard-core Bose gas with only one state per lattice site. We will 
solve the limit $N\to\infty$ within a saddle-point integration
and apply a
$1/N$-expansion to control systematically the corrections to the dilute Bose
gas towards a more strongly interacting bosons.

\section{BOSONS IN AN OPTICAL LATTICE}

In the first part of this section the Bose gas on a lattice will
be discussed without interaction using a complex field $\phi$. 
This field will be replaced in the second part by
a hard-core field which has the same propagator as the field
$\phi$. Therefore, in contrast to the Bose-Hubbard model,
the interaction is carried by the field and not directly by
the Hamiltonian.
A model of non-interacting bosons is considered on a
$d$-dimensional hypercubic lattice with ${\cal N}$ sites.
A boson can occupy statistically one of $N$ degenerate states at each site,
and tunneling between these $N$
states at lattice site $r$ to any of the states at site $r'$
occurs with the rate $J_{r,r'}^{\a,\a'}$.
This is represented by the Hamiltonian
\beq
{\hat H}=-{1\over 2}\sum_{\a,\a'=1}^N\sum_{r,r'}J_{r,r'}^{\a,\a'}
\ph^{+\a}_r\ph_{r'}^{\a'}
\label{hamilton}
\eeq
with bosonic creation and annihilation operators $\ph^{+\a}_r$,
$\ph_{r'}^{\a'}$.
The statistics of a system of bosons can be described by introducing
a grand-canonical ensemble with fugacity $\zeta$ at the inverse
temperature $\beta$. In the case of non-interacting bosons it is defined by
the partition function\cite{huang}
\beq
Z=\prod_j{1\over 1-\zeta^\beta e^{-\beta\epsilon_j}},
\label{part0}
\eeq
where $\epsilon_j$ are the eigenvalues of the Hamiltonian matrix
\[
H_{r,r'}^{\a,\a'}=-{1\over 2}J_{r,r'}^{\a,\a'}.
\]
Assuming that $\beta$ has integer values, the factors in Eq. (\ref{part0})
can be expressed by the determinant of a $\beta\times\beta$--matrix as
\[
1-\zeta^\beta e^{-\beta\epsilon_j}
=\det({\bf 1}-\zeta w_j),
\]
where
\beq
w_j=\pmatrix{
0 & u & 0 & \ldots & 0 \cr
0 & 0  & u & \ddots & \vdots \cr
\vdots & \ddots & \ddots & \ddots & 0 \cr
0      & \ldots & 0 & 0 & u \cr
u  & 0 & \ldots & 0 & 0 \cr
}\ \ \ {\rm with}\ \ u=e^{-\epsilon_j}.
\label{matrixw}
\eeq
Then the partition function can also be written as
the determinant of a $\beta{\cal N}\times\beta{\cal N}$--matrix as 
\beq
Z={1\over\det({\bf 1}-\zeta {\hat w})}.
\label{part01}
\eeq
${\hat w}$ is obtained from expression (\ref{matrixw}) by using
for $u$ the ${\cal N}\times{\cal N}$--matrix $u=e^{-H}$.
It is convenient for the following to introduce space-time coordinates
$x=(t,r)$ with a ``time'' variable $t=1,2,...,\beta$. Moreover, we assume
that the matrix $H=-\log u$ is chosen such that
\beq
{\hat w}_{x,x'}^{\a,\a'}={1\over N} w_{x,x'}.
\label{Ndep}
\eeq

\subsection{Random-Walk Expansion}

Using a complex field
$\Ph_x$ the inverse determinant in Eq. (\ref{part01}) reads
\beq
Z=\int \exp\Big[
-\sum_x\sum_{\a=1}^N\bph_{x}^\a\Ph_{x}^\a
+{\zeta\over N}\sum_{x,x'}\sum_{\a,\a'=1}^Nw_{x,x'}\bph_{x}^\a\Ph_{x'}^{\a'}
\Big]
\prod_x d\bph_x^\a d\Ph_x^\a/\pi.
\label{part00}
\eeq
The field is subject to periodic boundary conditions in $t$
\beq
\Ph_{\beta+1,r}^\a=\Ph_{1,r}^\a.
\label{pbc}
\eeq
The integrand of $Z$ in the expression (\ref{part00}) can be
formally expanded in powers of the matrix $w$ as
\beq
\exp\Big({\zeta\over N}\sum_{x,x'}\sum_{\a,\a'=1}^Nw_{x,x'}\bph_{x}^\a
\Ph_{x'}^{\a'}\Big)
=\prod_{t=1}^\beta\prod_{r,r'}\prod_{\a,\a'=1}^N
\Big(\sum_{l\ge0}{1\over l!}
[{\zeta\over N} u_{r,r'}\bph_{t,r}^\a\Ph_{t+1,r'}^{\a'}]^l\Big).
\label{he1}
\eeq
The integration over the Bose field $\phi_x$ can be performed for each
expansion term, leading to a random-walk expansion of 
$Z$ with elements $\zeta u_{r,r'}$.\cite{ziegler2,glimm}
From $Z$ we can evaluate the density of bosons as
\beq
n={\zeta\over{\cal N}N\beta}{\partial\over\partial \z}\log Z.
\label{totdens}
\eeq
In the case of the non-interacting Bose gas Eq. (\ref{part0}) we get
immediately
\[
n=-1+{1\over{\cal N}N\beta}{\rm Tr}[({\bf 1}-\zeta w/N)^{-1}]
={1\over{\cal N}N\beta}\sum_{l\ge1}\Big({\zeta\over N}\Big)^l
{\rm Tr} (w^l).
\]
$w^l$ contains $l-1$ summations with respect to $\a=1,...,N$,
contributing a factor $N^{l-1}$. Another factor $N$ comes from the
trace. Altogether, this gives $n\propto N^{-1}$,
a consequence of the scaling in Eq. (\ref{Ndep}).
It should be noticed that $\zeta$ is restricted here to $\zeta\le1$
if the eigenvalues of $w/N$ are $\le 1$.
This is a well-known artefact of non-interacting bosons\cite{huang}.
As we will see subsequently, $\zeta\ge0$ is not restricted in the
interacting Bose gas and has a non-zero density for $\zeta>1$ even in
the limit $N\to\infty$.

\subsection{The Hard-Core Bose Gas}

A hard-core condition can now be implemented in this model by assuming that
a crossing of different random walks is prohibited. For this purpose it is
useful to return to Eq. (\ref{he1}).
It has been shown\cite{ziegler2} that the statistics of the directed lines
with hard-core interaction can be described conveniently
by replacing the complex field $\Ph^\a_x$ by a field
constructed from an algebra of nilpotent numbers
(i.e., $(\eta_x^{\a,\s})^l=0$ if $l>1$):
\[
\Ph^\a_x\to\eta_x^{\alpha,\sigma}\ \ (\sigma=\pm1).
\]
Each space-time point $x$
is characterized by quantum numbers $(\a,\s)$,
where $\s=\pm1$ corresponds to
the real and the imaginary part of $\Ph^\a_x$. This choice
implies $2N$ degrees
of freedom at each point $x$. Therefore, we need also $2N$
variables
$\eta_x^{\a,\s}$ at $x$. Since $\{\eta_x^{\a,\s}\}$ are
nilpotent, we have
an exclusion principle describing the hard-core interaction. Products of
$\eta_x^{\a,\s}$ must be commutative in order to represent the Bose statistics.
Now one can identify an empty site with
\[
\eta_x^{\a,1}\eta_x^{\a,2}.
\]
and a tunneling process, going from $x$ to $x'$ and connecting the
states $\a$ and $\a'$, with
\[
{\zeta\over N} w_{x,x'}\eta_{x}^{\a,1}\eta_{x'}^{\a',2},
\]
where $w_{x,x'}$ ($=w_{x-x'}$) describes again
the hopping probability as obtained from the random-walk expansion.

The integration over the free Bose field $\phi$ 
is replaced by a linear mapping of the 
algebra, generated by $\{\eta_x^{\a,\s}\}$,
to the complex numbers with
$$
\int_{HC}
\prod_{x\in\L '\subseteq\L}\prod_{\a\in i_x}\prod_{\s\in j_{x,\a}}
\eta_x^{\a,\s}=\cases{
1& if $\L '= \L$, $i_x=1,2,...,N$, $j_{x,\a}=1,2$\cr
0& otherwise\cr
}.
$$
Thus the integral vanishes if the product is incomplete with respect to the
lattice, the degenerate states $\a$ or $\sigma$. Finally, 
we must impose periodic boundary conditions in the $t$-direction.
It is evident that we can introduce analytic functions of the nilpotent field.
Therefore, we may write for the partition function of the hard-core Bose gas
\[
Z_{HC}
=\int_{HC}\exp\Big[\sum_{x,x'}\sum_{\a,\a'=1}^N
(\delta_{x,x'}\delta_{\a,\a'}+{\zeta\over N} w_{x,x'})
\eta_{x}^{\a,1}\eta_{x'}^{\a',2}
\Big] .
\]
\no
The nilpotent field $\eta_x^{\a,\s}$ is closely related to a Grassmann field:
the algebra, generated by $\{\eta_x^{\a,\s}\}$, can be constructed
from anticommuting Grassmann variables\cite{NO} 
$\{ \psi_x^{\a,\s},\bpsi_x^{\a,\s}\}$ as
\[
\eta_x^{\a,\s}=(-1)^\s\psi_x^{\a,\s}\bpsi_x^{\a,\s} .
\]
Using the usual Grassmann integral\cite{NO}, the partition 
function now reads
\[
Z_{HC}=
\]
\beq
\int\exp\Big[\sum_{x,x'}\sum_{\a,\a'=1}^N
(\delta_{x,x'}\delta_{\a,\a'}+{\zeta\over N} w_{x,x'})
\psi_{x}^{\a,1}\bpsi_{x}^{\a,1}
\bpsi_{x'}^{\a',2}\psi_{x'}^{\a',2}\Big]
\prod d\psi_x^{\a,\s}d\bpsi_x^{\a,\s}.
\label{part3}
\eeq
Here we notice an identity for the diagonal term:
\[
\exp (\psi_x^{\a,1}\bpsi_x^{\a,1}\bpsi_{x}^{\a,2}\psi_x^{\a,2})=
1+\psi_x^{\a,1}\bpsi_x^{\a,1}\bpsi_{x}^{\a,2}\psi_x^{\a,2}
\]
\[
=\exp [(\psi_x^{\a,1}\psi_x^{\a,2}+\bpsi_x^{\a,1}\bpsi_x^{\a,2})]
-(\psi_x^{\a,1}\psi_x^{\a,2}+\bpsi_x^{\a,1}\bpsi_x^{\a,2}).
\]
The second term on the r.h.s. does not contribute to the partition function,
since there can only be a factor 
$u_{r,r'}\psi_{t,r}^{\a,1}\bpsi_{t,r}^{\a,1}$ or
$u_{r',r}\bpsi_{t,r}^{\a,2}\psi_{t,r}^{\a,2}$ from the bosons in 
(\ref{part3}). Thus,
the diagonal term in $Z_{HC}$ can be replaced by a term that is only
bilinear in the Grassmann field. The $w$--dependent term in $Z_{HC}$, which
describes the bosons, can also be expressed bilinearly in the
Grassmann field when we introduce two complex Gaussian fields (``Hubbard--
Stratonovich transformation''). Two fields are required here, since $w$ is
not a positive matrix. However, we obtain a positive matrix $v_s$ when we add
a positive diagonal matrix to $w$ as
$$
w_{x,x'}=w_{x,x'}+(s-s)\delta_{x,x'}=:(v_s)_{x,x'}-s\delta_{x,x'}, 
$$
with a sufficiently large $s$. (Physical results should not depend on $s$.
This will be confirmed in the results of the limit $N\to\infty$ below.)
With
$$
\rho_x=\sum_\a\psi_x^{\a,1}\bpsi_x^{\a,1},\ \ \brho_x=\sum_\a\bpsi_x^{\a,2}
\psi_x^{\a,2}
$$
we obtain the identity
$$
\Big({\pi \zeta^2\over sN^2}\Big)^{\beta{\cal N}}
\exp [{\zeta\over N}(\rho ,w\brho)]=
$$
\beq
\int\exp\Big[-{N\over\zeta}(\vphi,v_s^{-1}\bvphi)-
{N\over s\zeta}(\chi,\bchi)+(\rho,\bvphi)+(\vphi,\brho)+i(\rho,\bchi)+i(\chi,
\brho)\Big]\prod d\vphi_x d\chi_x,
\label{ident}
\eeq
where the scalar product means $(\vphi,v_s^{-1}\bvphi)=\sum_{x,x'}
\vphi_x v_{s;x,x'}^{-1}\bvphi_{x'}$. The r.h.s. of (\ref{ident}) can be
substituted into $Z_{HC}$, and the bilinear Grassmann term can be integrated.
This leads to
\begin{eqnarray}
Z_{HC} & = &
\Big({\pi \zeta^2\over sN^2}\Big)^{-\beta{\cal N}}
\int\exp\Big\{-N\Big[{1\over\zeta}(\vphi,v_s^{-1}\bvphi)+{1\over s\zeta}
(\chi,\bchi)
\nonumber\\
&-&\sum_x\log [1 +(\vphi_x+i\chi_x)(\bvphi_x+i\bchi_x)]\Big]\Big\}\prod
d\vphi_x d\chi_x.
\label{part2}
\end{eqnarray}
It should be noticed that the field $\chi$ can also be integrated
out in principle. However, this is difficult if $N\gg1$.

We notice that the expression in the exponent is invariant under a global
phase transformation of the complex field: $\vphi$, $\chi$ $\to$ $e^{i\kappa}
\vphi$, $e^{i\kappa}\chi$ and $\bvphi$, $\bchi$ $\to$ $e^{-i\kappa}
\bvphi$, $e^{-i\kappa}\bchi$. Furthermore, the logarithmic term of the
action is also symmetric under a local $U(1)$-transformation. This means that
the depencence of the logarithmic term on the local phase of $\vphi_x$
can be gauged away, and this term depends only on the modulus of $\vphi_x$. 
We will see in Sect. 5 that the symmetry under
this phase transformation is spontaneously broken. As a
consequence, there is a one-component Goldstone mode.

The total density of bosons can be expressed within the new
effective complex fields $\varphi$ and $\chi$.
From Eq. (\ref{totdens}) we obtain
\beq
n={\zeta\over N{\cal N}\beta}{\partial\over\partial\zeta}\log Z_{HC}
={1\over \zeta{\cal N}\beta}\Big[\langle(\varphi,v_s^{-1}{\bar\varphi})
\rangle+{1\over s}\langle(\chi,{\bar\chi})\rangle\Big]+{2\over N},
\label{totdens1}
\eeq 
where
\begin{eqnarray}
\langle ...\rangle & = &
\Big({\pi \zeta^2\over sN^2}\Big)^{-\beta{\cal N}}{1\over Z_{HC}}
\int ... \exp\Big\{-N\Big[{1\over\zeta}(\vphi,v_s^{-1}\bvphi)+{1\over s\zeta}
(\chi,\bchi)
\nonumber\\
&-&\sum_x\log [1 +(\vphi_x+i\chi_x)(\bvphi_x+i\bchi_x)]\Big]\Big\}\prod
d\vphi_x d\chi_x.
\end{eqnarray}

\section{$N\to\infty$: CLASSICAL-FIELD EQUATIONS}

The action in (\ref{part2}) depends on the number of states only
through the
prefactor $N$. This suggests that a saddle-point integration can be
performed for $Z_{HC}$. For this purpose it is convenient
to rescale the fields first:
\[
\varphi\to \zeta^{-1/2}\varphi,\ \ \ \chi\to \zeta^{-1/2}\chi.
\]
We will use the same name for the rescaled fields. Then the 
saddle-point equations read
\beq
{\partial S\over\partial\vphi_x}=\sum_{x'}(v_s^{-1})_{x,x'}\bvphi_{x'}-
{\bvphi_x+i\bchi_x\over \z^{-1}+(\vphi_x+i\chi_x)(\bvphi_x+i\bchi_x)}=0 
\label{spa1}
\eeq
\beq
{\partial S\over\partial\bvphi_x}=\sum_{x'}(v_s^{-1})_{x',x}\vphi_{x'}-
{\vphi_x+i\chi_x\over \z^{-1}+ (\vphi_x+i\chi_x)(\bvphi_x+i\bchi_x)}=0
\label{spa2}
\eeq
\beq
{\partial S\over\partial\chi_x}={1\over s}\bchi_x-
{i(\bvphi_x+i\bchi_x)\over \z^{-1}+ (\vphi_x+i\chi_x)(\bvphi_x+i\bchi_x)}=0
\label{spa3}
\eeq
\beq
{\partial S\over\partial\bchi_x}={1\over s}\chi_x -
{i(\vphi_x+i\chi_x)\over \z^{-1}+ (\vphi_x+i\chi_x)(\bvphi_x+i\bchi_x)}=0
\label{spa4}
\eeq
Eqs. (\ref{spa1}) - (\ref{spa4}) correspond to a discrete version of
the Gross-Pitaevskii (or non-linear Schr\"odinger)
equation, often used for the description of a weakly-interacting
Bose gas\cite{ginzburg,baym}, a case in which we can assume that
fields are small. Expansion in terms of the fields yields an equation for
$\vphi$:
\[
\sum_{x'}(v_s)^{-1}_{x',x}\vphi_{x'}
+\Big[-{1\over s+z}+{z^2\over (s+z)^4}\vphi_x\bvphi_x
\Big]\vphi_x\sim 0.
\]
In general, it is not possible
to solve these equations. Therefore, the assumption of a
uniform solution is useful. This is known as the Thomas-Fermi
approximation\cite{baym}.

\subsection{The Thomas-Fermi Approximation}

For a uniform solution we obtain with
$\sum_{x'}(v_s^{-1})_{x,x'}=1/(1+s)$
and Eqs. (\ref{spa1})-(\ref{spa4})
\[
i\chi =-{s\over 1+s}\vphi ,\ \ \ \ i\bchi =-{s\over 1+s}\bvphi .
\]
There is always a trivial solution 
$\vphi_0 =\bvphi_0=\chi_0=\bchi_0=0$ and a
non--trivial solution with
\[
\vphi_1\bvphi_1=(1+s)^2(1-z).
\]
The fugacity $\zeta$ has been replaced here by the inverse
fugacity $z=\z^{-1}$.
The saddle-point contribution from the trivial solution
gives for the partition
function $Z_0=1$.
Any non--trivial solution breaks the symmetry under phase transformation for
$z\ne 1$. The saddle-point
equations, however, are invariant under the symmetry transformation because
the phase factor is not determined by them. This leads to massless
fluctuations that generate long-range correlations. 
The $N\to\infty$ limit of the partition function is not affected
by these fluctuations
\[
Z_1=z^{-N{\cal N}\beta}e^{(z-1)N{\cal N}\beta}.
\]
Since the free energy $\ -\log Z_{HC}$ must
be minimal, the non--trivial saddle point with 
$Z_{HC}=Z_1$ is valid for $z<1$ whereas the
trivial saddle-point solution with $Z_{HC}=Z_0$ is valid for
$z>1$. This implies for the density of bosons of
Eq. (\ref{totdens1}) in the limit $N\to\infty$
\[
n(z )\to\cases{
1-z & for $z<1$\cr
0 & for $z>1$\cr
}.
\]
Thus $z=1$ is the critical point. According to the definition of
$z$ as the inverse fugacity of the bosons, $z=1$ separates the
condensed phase ($z<1$) from the normal-fluid phase ($z>1$).
Corrections of $o(N^{-1})$ to $n(z)$ will be evaluated in Sect. 5.1.

\section{$1/N$-CORRECTIONS: QUASIPARTICLES}

In order to investigate the stability of the saddle-point
solutions against fluctuations, we shall consider now
$$
\vphi\approx\vphi_1+\dphi'+i\dphi'',{\ \ }
\chi\approx\chi_1+\dchi'+i\dchi''.
$$
With $\Dphi=(\dphi',\dphi'',\dchi',\dchi'')$
we obtain for the partition function in Gaussian approximation
\beq
Z_{HC}\approx Z_\nu\int\exp [-N(\Dphi,\cG_{s,\nu}^{-1}\Dphi)]\prod_x
d\Dphi_{x}
\label{part4}
\eeq
with the propagator (Green's function) $\cG_{s,\nu}$.
It remains to evaluate the latter.  
For this purpose it is necessary to choose a specific $u=e^{-H}$ 
(or $H=-\log u$), for instance,
\beq
u_{r,r'}=\cases{
J/2d & for $r,r'$ nearest neighbors \cr
(1-J) & for $r'=r$ \cr
0 & otherwise \cr
}.
\label{w}
\eeq
Then the eigenvalues of the matrix $w$ are
\beq
{\tilde w}=e^{i\omega}[1-J+{J\over d}\sum_{j=1}^d\cos(k_j)]
\label{eigenvalue}
\eeq
with Matsubara frequency $\omega$ and $d$-dimensional wavevector components
$k_j$ with
$-\pi\le \om, k_j <\pi$. 
The eigenvalues are bounded by $|{\tilde w}|\le1$.
It is sufficient to consider $s=1$ to get a positive matrix $v_1$.
Then a simple calculation gives the Green's function
$$
\cG_{1,\nu}^{-1}\equiv\cG_\nu^{-1}=\pmatrix{
{\hat v}_1^{-1}-A_\nu&-iA_\nu\cr
-iA_\nu&{\bf 1}+A_\nu\cr
}
$$
with
\[
{\hat v}_1^{-1}
\sim{1\over2}\pmatrix{
1+k^2/4\tau & -\om/2\cr
-\om/2&1+k^2/4\tau\cr
}
\]
for small wavevectors $k$ and frequencies $\omega$. 
$\nu=0,1$ refers to the two saddle-point solutions within the
Thomas-Fermi approximation. The new 
parameter $\tau$ is proportional to the inverse tunneling rate on the
$d$-dimensional lattice: $\tau=d/J$.
The $2\times2$ block matrices depend on the fugacity $1/z$
\[
A_0={1\over z}\pmatrix{
1&0\cr
0&1\cr
},\ \ \ A_1=\pmatrix{
2z-1&0\cr
0&1\cr
}.
\]

\section{RESULTS}

The spectrum of the propagator of the field $\varphi$ is given
as (s. App. A)
\beq
\eps(k)=\sqrt{{\zeta-1\over\tau}k^2+({k^2\over 2\tau})^2}.
\label{spectrum}
\eeq
For a weakly-interacting Boge gas of bosons with mass $m$ there is the 
well-known Bogoliubov spectrum\cite{popov}
$$
\eps(k)=\sqrt{\mu_0 k^2/m+(k^2/2m)^2}.
$$
Since $\mu_0=\log\z $ we can identify our parameter
$\tau$ with the mass of weakly-interacting bosons. Then
Eq. (\ref{spectrum}) agrees with the Bogoliubov spectrum only
for $\z \sim1$, indicating that the dilute limit of the hard-core Bose
gas is a weakly-interacting system.
The Goldstone mode of $\cG_1$ is proportional to
$$
{\zeta -1\over\tau}k^2-\om^2.
$$
Thus, the sound velocity is $c=\sqrt{(\zeta -1)/\tau}$. On the
other hand, Popov\cite{popov} finds $c=\sqrt{\mu_0/m}$.
Therefore, the sound velocity of the $N$ state hard-core Bose gas agrees
with that of the weakly-interacting bosons only for $\z\sim1$.

\subsection{Total Density of Bosons}

From Eq. (\ref{totdens1}) we obtain the expression
\beq
n=(1-z)\Theta(1-z)+n_1/N +o(N^{-2})
\label{totdens3}
\eeq
with 
\[
n_1=2+{1\over \beta{\cal N}}{\rm Tr}({\cal G}_\nu{\partial\over\partial\zeta}
{\cal G}_\nu^{-1}).
\] 
Here $\Theta(x)$ is the Heaviside step function.
This result indicates that there are no bosons at $z>1$ in the
limit $N\to\infty$, as in the non-interacting Bose gas. However,
the density increases linearly for $z<1$, a regime that is not accessible
for the non-interacting Bose gas.
For a finite number of states $N$ there is a non-vanishing
density of particles for any value of the inverse fugacity $z$.

\subsection{Density of the Condensate}

The condensate can be studied by considering the correlation function
of the hard-core Bose field
\[
C_{x,x'}={1\over N^2}\sum_{\alpha,\alpha'=1}^N
\langle\eta_x^{\alpha,1}\eta_{x'}^{\alpha,2}\rangle.
\]
On large scales this does not decay in the condensed phase but exhibits
off-diagonal long range order:
\[
C_{x,x'}\sim n_0\ \ \ (|r-r'|\sim\infty),
\]
where $n_0$ is the density of the condensate. Using the notation of
Sect. 2 we can also express the correlation function of the Bose field as
\[
C_{x,x'}={1\over N^2}\langle\rho_x{\bar\rho}_{x'}\rangle.
\]
\begin{figure}
\begin{center}
\mbox{\psbox{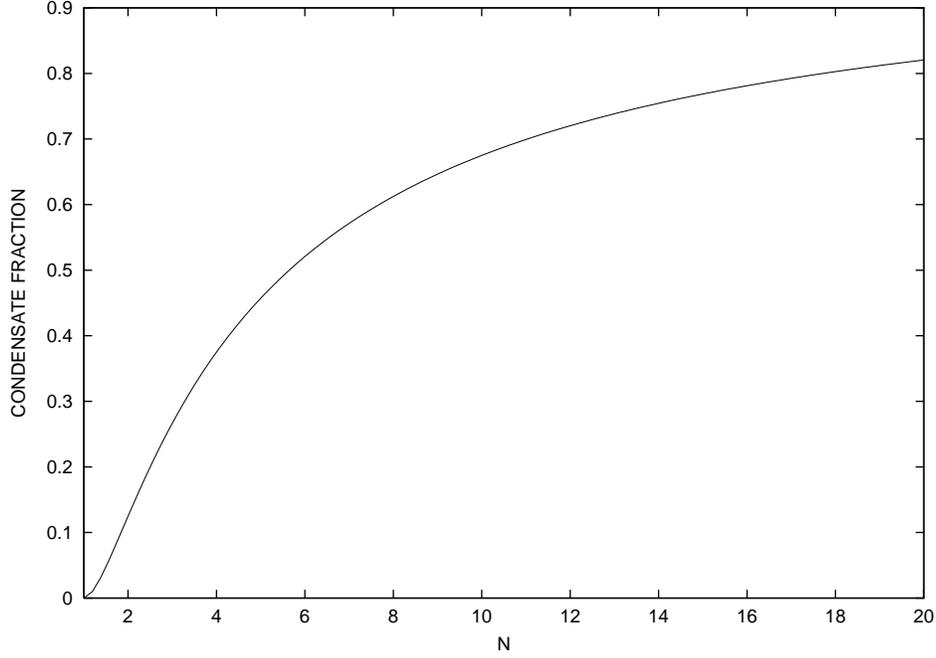}}
\caption{Fraction of the condensate $n_0/n$ as a function of 
the number of degenerate states $N$ per site at fugacity $\zeta=2$
(from Eq. (\ref{fraction})).}
\end{center}
\end{figure}
Eventually, in terms of the fields $\varphi$ and $\chi$ the latter reads
\[
C_{x,x'}=-{1\over N}(v_s^{-1})_{x',x}+
\]
\[
\Big\langle
\Big[\vphi_x'-{1\over N}{\varphi_x+i\chi_x\over
z+(\varphi_x+i\chi_x)({\bar\varphi}_x+i{\bar\chi}_x)}\Big]
\Big[\bvphi_{x'}'-{1\over N}{{\bar\varphi}_{x'}+i{\bar\chi}_{x'}\over
z+(\varphi_{x'}+i\chi_{x'})({\bar\varphi}_{x'}+i{\bar\chi}_{x'})}\Big]
\Big\rangle
\]
with $\vphi_x'=\sum_{x'}(v_s^{-1})_{x,x'}\vphi_{x'}$.
Using the result of the saddle-point integration this expression leads
for $|r-r'|\to\infty$ to
\beq
n_0=(1-{1\over N})^2(1-z)
\Theta(1-z)
\label{cond1}
\eeq
for the density of the condensate. The prefactor $(1-1/N)^2$
indicates a suppression of the condensate
due to increasing interaction with a decreasing number of particles
states $N$ at each lattice site. In Eq. (\ref{cond1}) terms of higher
order in $1/N$ are neglected. It is expected that they lead to additional
terms such that $n_0$ does not vanish at $N=1$.
Combining (\ref{totdens3}) and (\ref{cond1}) the condensate fraction reads
\beq
{n_0\over n}=(1-{1\over N})^2{(1-z)\over 1-z+n_1/N}\Theta(1-z)
\label{fraction}
\eeq
which is plotted in Fig.1.

\section{CONCLUSIONS}

A hard-core Bose gas on a lattice has been used to study the effect
of interaction in an optical lattice. Introducing $N$ degenerate states at
each lattice site and a hard-core interaction only between the same state,
the limit $N\to\infty$ can be solved by a saddle-point integration.
This limit exhibits a transition from a normal state to a Bose-Einstein
condensate if the fugacity $\zeta=1$. The limit $N\to\infty$ is very
special, since the total density of bosons vanishes for $\zeta<1$.
However, $1/N$-corrections lead to a non-vanishing density of bosons
for any value of $\zeta$. On the other hand, $1/N$-corrections indicate
a depletion of the condensate ($\zeta>1$). These results agree with
the observation that the condensate fraction decreases with an increasing
coupling constant in the Bose-Hubbard model\cite{oosten}. Further studies
are neccessary in order to find the effect of very strong interaction
when $N\sim1$.
\vskip4mm

\section*{ACKNOWLEDGMENTS}

 The author would like to thank T. Esslinger for
discussing recent experiments with Bose-Einstein condensates in
optical lattices.

\section*{APPENDIX A: EFFECTIVE PROPAGATOR FOR THE FIELD $\varphi$}

Starting from the partition function $Z_{HC}$ of Eq. 
(\ref{part4}), the field $\chi$ can be integrated
out. The result is an effective propagator of the field $\varphi$
\[
{\cal G}_\nu=(v_1^{-1}-A_\nu+B_\nu)^{-1},
\]
where we have
\[
B_0={1\over z(1+z)}\pmatrix{
1&0\cr
0&1\cr
},\ \ \ B_1={1\over2}\pmatrix{
(2z-1)^2/z & 0 \cr
0 & 1 \cr
}.
\]
Then a straightforward calculation gives the following
expressions
\[
{\cal G}^{-1}_0\sim{1\over2}\pmatrix{
(z-1)/(z+1)+k^2/4\tau & -\om/2 \cr
-\om/2 & (z-1)/(z+1)+k^2/4\tau \cr
}
\]
and
\[
\cG^{-1}_1\sim{1\over2}\pmatrix{
(1-z )/z +k^2/4\tau & -\om/2\cr
-\om/2 & k^2/4\tau \cr
}.
\]
From $\cG^{-1}_1$ we obtain for the spectrum of the condensed
phase (i.e. $z<1$) the expression
$$
\eps(k)=\sqrt{{(1-z )\over z }{k^2\over\tau}+({k^2\over 2\tau})^2}.
$$

\end{document}